\documentclass[10pt]{article}
\usepackage{atbegshi}
\AtBeginDocument{\AtBeginShipoutNext{\AtBeginShipoutDiscard}}
\usepackage{graphicx}
\usepackage{amsmath}
\usepackage{amsfonts}
\usepackage{amssymb}
\usepackage{float} 
\begin{document}
\begin{center}
\author{Becker, F. M.
\and 
Bhatt, Ankur S.}
\end{center}
\title{Electrostatic accelerated electrons within symmetric capacitors during field emission condition events exert bidirectional propellant-less thrust}
\maketitle

\begin{abstract}
During internal discharge (electrical breakdown by field emission transmission) thin symmetric capacitors accelerate slightly towards the anode; an anomaly that does not appear obvious using standard physics. Various thicknesses of discharging capacitors have been used to demonstrate and better characterize this phenomenon. It was observed that it is possible to reverse the force by adding conductive materials in the immediate proximity of the cathode when physically separated from the anode (thus not galvanically connected). Conversely, the addition of conductive materials in the area surrounding the anode did not alter the original force observed. The data gathered seems to confirm a phenomenon that could be exploited for propulsion purposes, in particular for fuel-less applications in a vacuum. The results could be correlated to an external cause which appear to be influenced by the particles' acceleration. Overall, the preliminary results are encouraging for practical engineering purposes.
\\

\end{abstract}
\setcounter{page}{1}

\section{Introduction}
\parindent0pt
It was experimentally observed (Becker F.M. 1990) that a thinly charged parallel plate capacitor, supplied with high voltage values of $5$ kV DC - $10$ kV DC, exerted an unexpected observable force towards the anode. This anomaly was only detected while using dielectrics with low dielectric breakdown strength. However, it progressively disappeared when increasing the dielectric breakdown voltage (and the material thickness). Such phenomenon was initially disregarded as a likely artifact, but a few years later, additional considerations led to the speculation that electric discharge could have been the cause of the anomaly. However, no further testing was conducted. The initial observations suggested that stronger dielectrics, with their enhanced performance in withstanding voltage, would not lead to an observable effect while the electric discharge (breakdown of insulation leading to partial/full discharge or field emission) could be responsible for the appearance of the phenomenon. In addition, Talley R.L. \cite{ll} had described a comparable anomalous observation which might have been caused by accelerating electrons or electric charges, further reducing the likelihood of a simple artifact.
\\

While conducting a variety of capacitive discharge experiments in 2017/2018 and additional battery powered wireless experiments in 2019, data was collected during field emission and insulation breakdown discharge events of parallel capacitive charged plates. Data was collected for very short capacitor plate distances. Additionally, the collected data investigated any correlation between any anomalous force and the accelerated mass and any anomalous force and the decrease of the capacitor electrode distance, while keeping the accelerated mass constant. 
\section{Method}

{\large\bf{Experimental setup and test apparatus}}

\subsection{}
Electrons are an easy-to-control option to achieve high acceleration of particles. Also with reference to the anomalies characterized by Talley R.L. \cite{ll}, such conditions are observed as potentially relevant. The high field strength achieved in the thin electrode separation can provide constant electron accelerations in the magnitude of $10^{19}$ [m s$^{-2}$] (particle acceleration equals the fundamental electric charge divided by mass of an electron multiplied with voltage divided by the electrode distance). The strong electric field releases electrons through the field emission effect \cite{ff}. With the objective of achieving higher force values, the material was heated so that the energy value required by the electric field for electron transmission could be lowered \cite{ii} (Schottky effect): in other words, warming the material allowed an increase in the discharge of electrons using the same strength of the electric field. The application of heat enables thermionic emission as well as supplying some electrons with at least the minimal energy required to overcome the barrier force holding them in the material structure (reference to the concept of work functions of materials).
\\
\subsection{}

Capacitors were designed with polyethylene dielectric materials. Other materials, such as paper, glycerin, and porous plastics, were also tested but this resulted in discharges involving ionic charge flow which traveled in the opposite direction of the electron flow hindering the effect. Furthermore, some materials are prone to cavities (atmospheric voids) that introduce partial discharges \cite{dd} where electron avalanches generate ion current contribution inside the hollow space. Additionally, paper insulators were ineffective in generating measurable thrust effect since the atmospheric electron avalanches (see Paschen law \cite{cc} for breakdown voltage vs distance) would release secondary electrons but also induce ionization as well. Overall, the experimental data seem to validate that, for the design and manufacturing of a thruster device using the observed phenomenon, a vacuum propagation of the accelerated electrons, or the application of semiconductor cathode arrays, could be the most effective method to obtain a controllable effect. 
\begin{figure}[H]
\centering
\includegraphics[scale=.75]{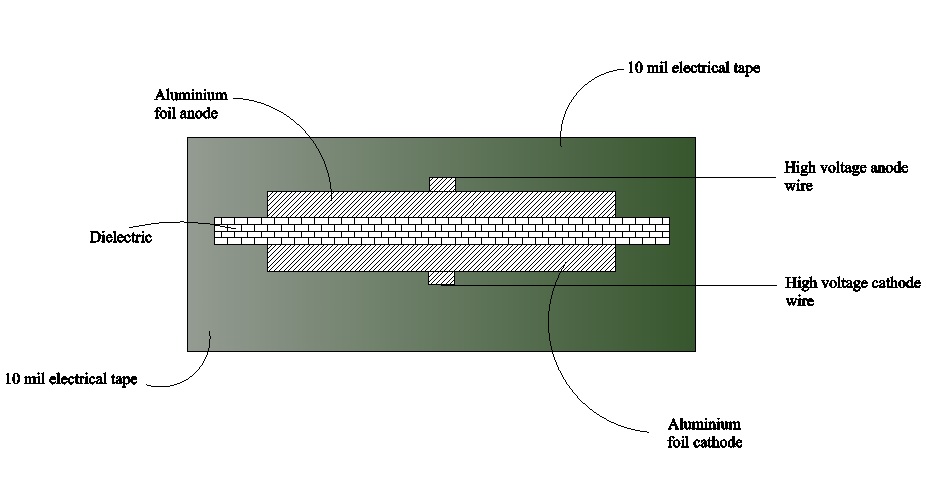}
\small{\caption{General prototype construction}}
\end{figure}
Remark: Depending on the actual insulator thickness (as described in this document) it can be assumed that currents slightly below 1 $\mu$A are essential to generate thrust forces to trigger the load sensor (i.e. LEADZM B300T digital scale with typical sensitivity of 0.001 mg and error of 0.003 mg).
\\

Considering the available power sources ($5$ kV DC, $10$ kV DC), the electric field strength between the capacitor plates would be estimated to be below the actual effective tunneling field strength required to provide a sufficiently reliable current density to support the effect. Therefore, the homogeneous field character had to be enhanced into a partial inhomogeneous field to increase the field strength by altering the smooth flat shape of the emitting surface and adding sharper edges (reduced radius of an emitting surface corresponds to increase in the electric field strength as similar to a concept of a needle cathode). This was done by cutting into the electrode or by using sandpaper on the cathode surface to facilitate field emission. This was done using a precision knife applying a high number (hundreds) of small cuts on the electrode or by surface treatment with a fine sandpaper. Obtaining a sharp edge contributes to a higher field strength through the creation of inhomogeneous electric fields, in analogy with a needle electrode. If only homogenous fields are used, this (cold) emission would need an electric field strength to begin on the order around 10$^7$ to 10$^8$ [V/m].
\\

For some tests, typically associated with higher thrust force values, the overall capacitor was preheated for testing up to approximately $50$ $^\circ$C before placing the device onto the measurement apparatus. This also denotes that testing attempts at low ambient temperatures without pre-heating may lead to a thrust force too low to be detected.
\\
\subsection{}
Confirmation tests in soft vacuum were conducted by placing the capacitor inside a sealed container, shielded by wrapping it in a conductive (grounded) outer layer. Such tests yielded results comparable, in terms of average, with the set-up of open air. Nonetheless the corresponding spread standard deviation of the data points recorded was lower for the soft vacuum ($\sim$20 torr) during experimental tests. 
\\
\subsection{}
The supply wires had been twisted \cite{bb} to reduce electromagnetic effects (Lorentz force etc.). This operation was performed very carefully to prevent the generation of torque onto the system. Nevertheless, the order of magnitude of theoretical torque contribution was estimated to be sufficiently low for not altering the observable phenomenon, reducing the concerns related to any possible residual torque. Moreover, the supply conductors were routed to an adequate distance ($\sim$200 mm) to prevent electromagnetic field disturbances on the load cell of the digital scale. Field disturbances to the load cell leading to the corruption of displayed measurements were observed in the range less than $\sim$50 mm around the digital scale. The metal film shunt resistors used for voltage measurements to determine the electric current had a 1 $\%$ error tolerance while the probes used were approximately 2 $\%$. The scope used for electric current measurements was a Keysight DSO1052B Oscilloscope-2 Channel-50 MHz.
\begin{figure}[H]
\begin{center}
\includegraphics[scale=0.45]{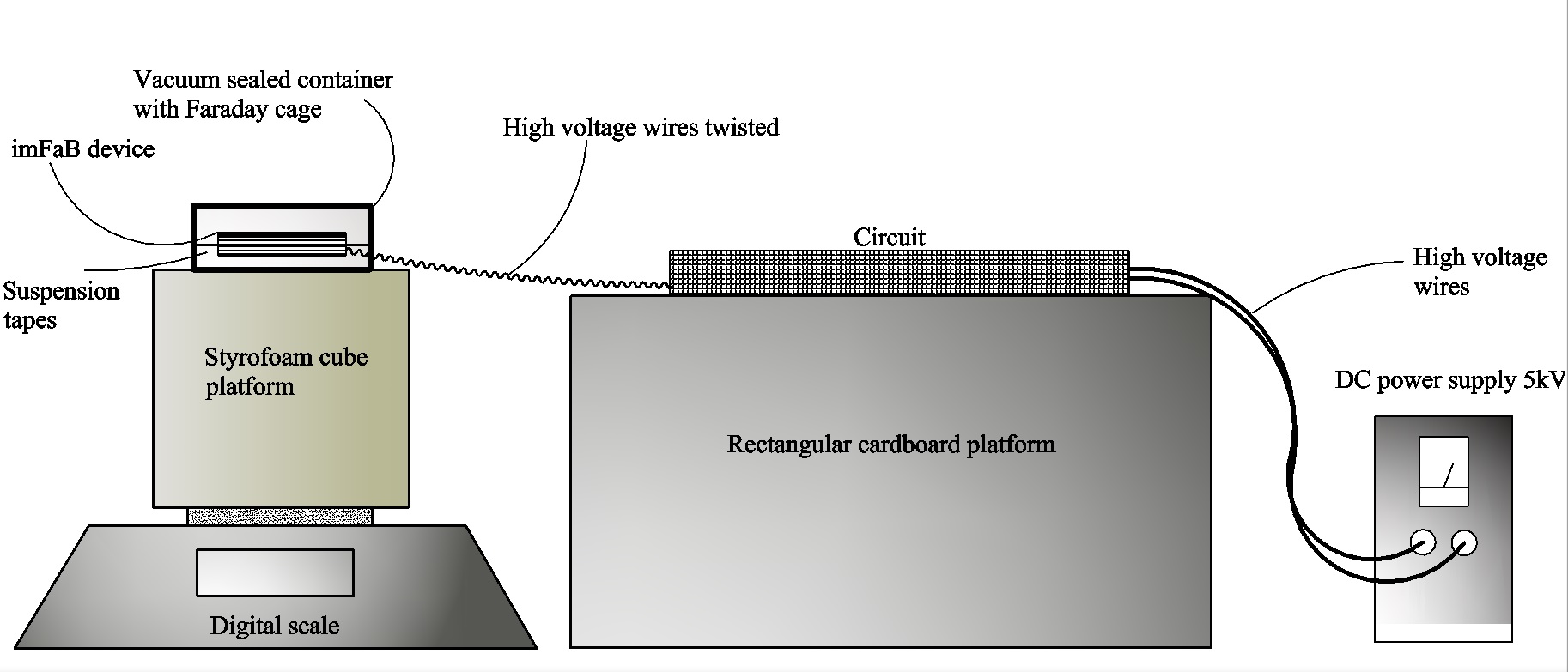}
\small{\caption{Experiment setup}}
\end{center}
\end{figure}
\subsection{}
The dielectrics had been tightly fit between the conductive surfaces to minimize the air presence. If a considerable air layer is present, in fact, this could lead to ions traveling in the opposite direction of the electrons due to avalanche processes by direct arcing, dampening the amplitude of observed thrust. As previously explained, this might also occur in certain materials when the high-field strength introduces partial internal discharges in material porosites where electrons as well as secondary avalanche electrons can generate positive charged ions. Furthermore, when glow-discharge/stronger arcing occurs, the acceleration voltage over the capacitor would decrease significantly, down to a typical $\sim$30V typical arc voltage level. In this condition particles, instead of being linear accelerated, would be subject to Langmuir waves, thus (plasma) rapid oscillations of the electron density related to the instability in the dielectric function. The experiment was consequently set up to minimize the risk of these occurrences (which sometimes had been observed with the test device accelerating slightly towards the earth regardless of the polarity). 
\\
\subsection{}
It was essential to protect the measurement tools/devices of the circuit from the influences of voltage transients that could occur in a situation where the insulation resistance of the capacitor would drop significantly. In such cases, the shunt resistor would carry a higher voltage due to the voltage divider characteristic of the circuit (more precisely, this is related to the capacitor insulator experiencing a reduced resistance). The circuit, see Fig. 4, does not employ any preventive measures for such phenomena, with regards to the application of transient surge protective devices. However, a large resistor (compared to the shunt resistive value) in series with the shunt could be introduced to serve as a voltage divider in case of insulation failure. Such transient effects are seen more likely when the field emission current increases. In addition, heating the dielectric insulation material could potentially reduce the insulation resistance of the dielectrics.
\\
\subsection{}
An alternative to the $5$ kV DC commercial power source was also used. A conventional flyback transformer with a maximum output slightly above $10$ kV DC has been utilized for voltage sweeping during the experiments to vary the electron's acceleration. The flyback transformer was supplied in the primary winding by a conventional DC power supply. The output voltage characteristic has been interpolated by the input/output data specifications of the flyback manufacturer for computations. Additionally, the electric circuit was rectified (smoothed) by a tank capacitor to provide a stable supply voltage to the capacitor. 
\begin{figure}[H]
\begin{center}
\includegraphics[scale=1]{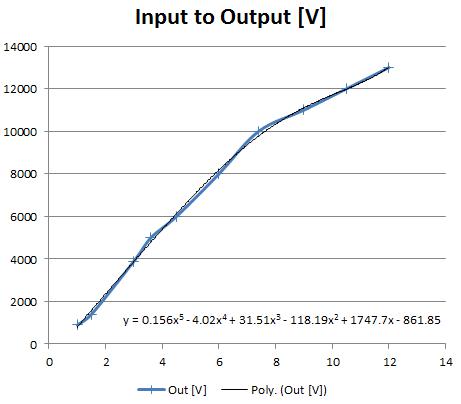}
\small{\caption{Interpolated output function for flyback supply source}}
\end{center}
\end{figure}
\begin{figure}[H]
\begin{center}
\includegraphics[scale=0.45]{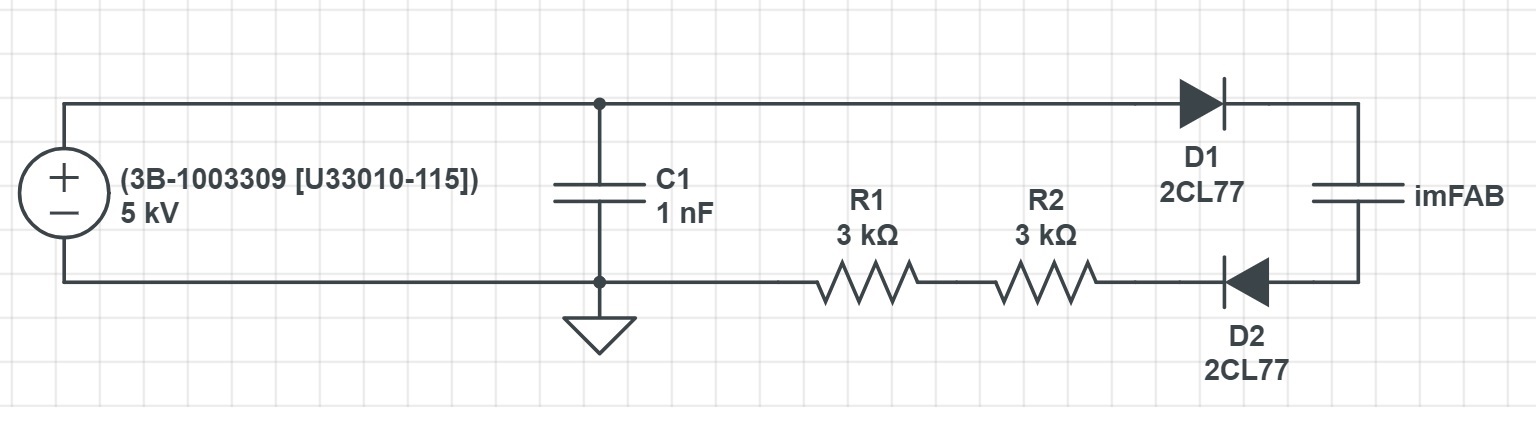}
\small{\caption{Functional circuit}}
\end{center}
\end{figure}

\subsection{Wireless Setup}

During additional testing in January 2019, the entire thruster system was placed inside a grounded faraday cage with an inner insulation barrier. Only DC supply wires exited the cage and were connected to a battery-onboard. These low voltage DC supply wires ran vertically and twisted until terminal connections at the flyback transformer.  All high voltage components were located inside the cage without any atmospheric exposed HV supply conductors. The thruster box, including the battery power supply and high voltage source, was placed on the scale and the center of gravity effects were addressed before measurement. Additionally, a wireless relay was installed on the top of the cage to turn the system on and off remotely without physical access. Signal conductors were routed to the measurement circuit (shunt to the oscilloscope). The effect device ON voltage level was fixed to around ~5kV  which was measured from the output of the flyback transformer with smoothing capacitor in parallel.

\begin{figure}[H]
\begin{center}
\includegraphics[scale=0.45]{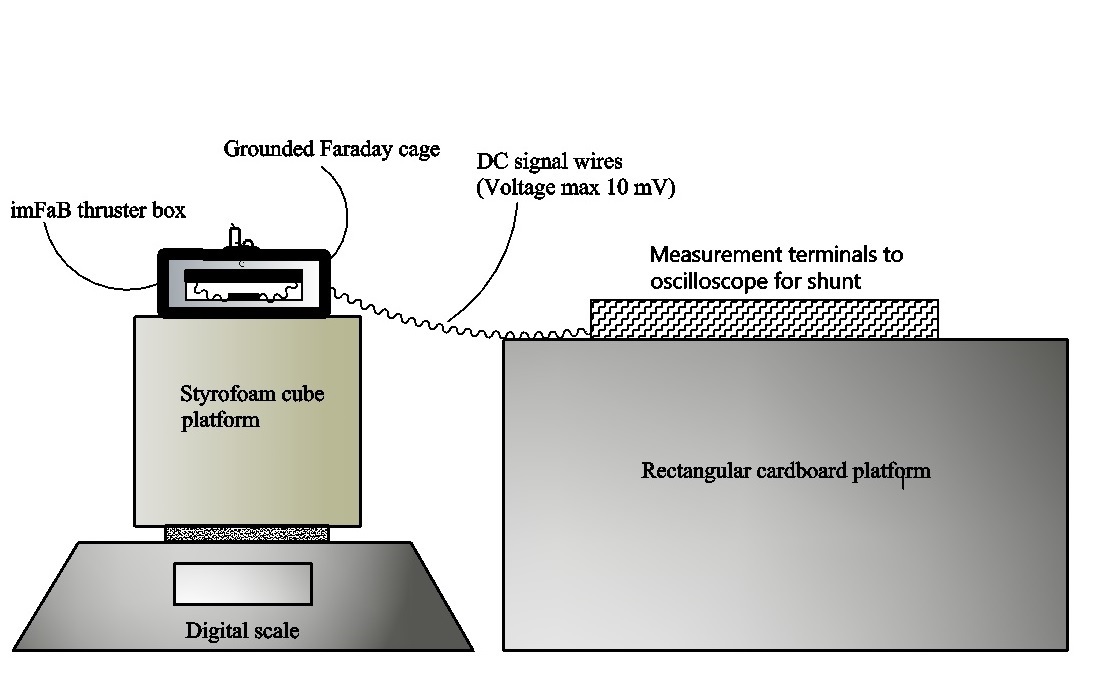}
\small{\caption{Wireless setup}}
\end{center}
\end{figure}

\begin{figure}[H]
\begin{center}
\includegraphics[scale=0.6]{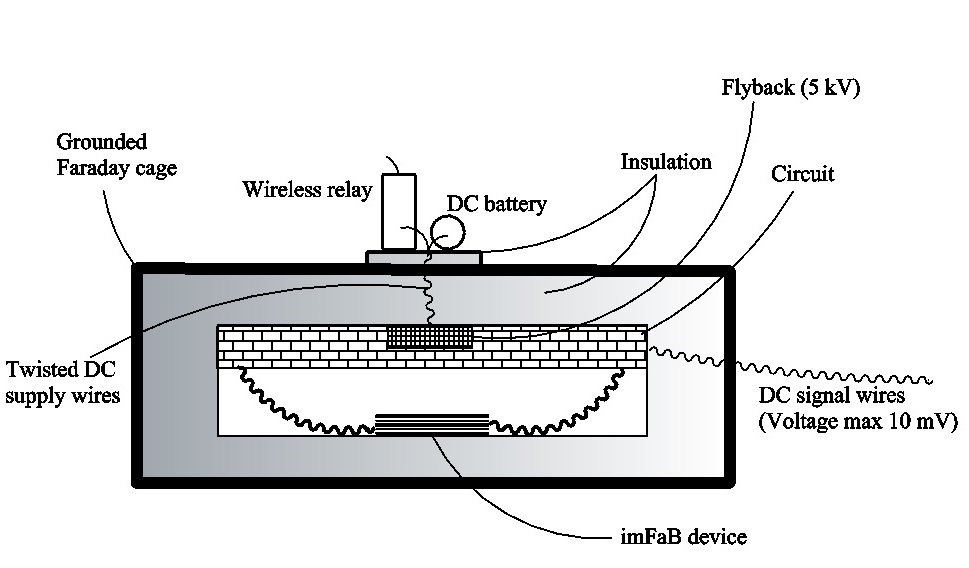}
\small{\caption{Thruster device}}
\end{center}
\end{figure}

\section{Results}
\subsection{Thrust observation versus electrode distances}

The electrode distance had been varied and the force monitored. Since the discharge current is dependent on the Schottky effect (field enhanced thermionic emission), which is determined not only by the electric field strength but also by the heat of the surface, the actual current for each data point statistically fluctuates. Consequently the graphs illustrating the results of the current measurements are normalized to a constant set-up (same charge amount, same accelerated mass).
\\

If a constant current (accelerated mass of electrons) is provided, the reduction of the distance between the electrodes causes the thrust force effect to exponentially increase in value. The electrodes distance is equivalent to the dielectric insulation thickness, and in the remainder of the paper the two dimensional characteristics are used interchangeably being equal in value.
\\

The first sub-experiment was conducted with different insulator thicknesses, between $13$ $\mu$m to $80$ $\mu$m, which influenced the acceleration experienced by the electrons. The diameters of the capacitors ranged from $2.5$ $\mu$m to $5$ $\mu$m and a total of $266$ data points were collected. The measured current value was standardized (as fluctuating under the effect of field emission) to a normalized current level of an arbitrary $10$ $\mu$A. The accelerated voltage was tuned to $5$ kV for all measurement points while testing different insulation thicknesses. The acceleration voltage of approximately $5$ kV corresponds to an electric acceleration inside the insulator thicknesses on the order of $10^{18}$ to $10^{19}$ [m s$^{-2}$].
\\

While altering the electrode's distance, it is observed that the recorded force follows a linear trend in a logarithmic view when the current is stablized to a normalized value. Hence the trend is attributed to an exponential increase of force by a linear reduction of the electrodes distance. 
\\

\begin{figure}[H]
\centering
\includegraphics[scale=1.3]{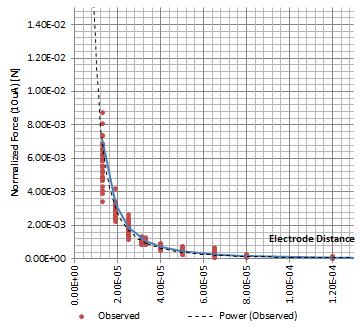}
\small{\caption{Normalized Force versus electrode distance}}
\end{figure}
\begin{figure}[H]
\centering
\includegraphics[scale=1.3]{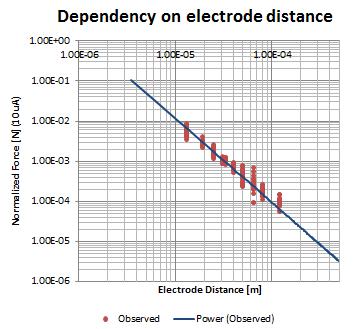}
\small{\caption{Normalized Force versus electrode distance (log)}}
\end{figure}

Unless otherwise specified, data points of individual graphs are considered an absolute value since these points correspond to both directional scenarios (force towards measurement scale and opposite). The anode of the capacitor pointing upwards was associated to an upward acceleration while the anode pointing downwards was associated to a downward acceleration. This remained valid as long as no reversal force mechanism was introduced (see section 3.3 for details).
\\

In Fig. 9 the actual force shows dependency on the accelerated electron mass indirectly measured by the value of the corresponding electric current. The symmetry of the data points illustrates that the observed force appears independent from influences of gravity, ionic winds and thermal buoyancy (for the latter, see additional clarifications in following paragraphs).
\\

Additionally, several tests have been conducted in low vacuum using a simple sealed container (data points identified as ‘vac’ in the attached graph), showing a trend line characterized by a lower standard deviation compared to the open-air tests.
\\

Note: Vacuum measurement of the median thrust observations (conducted on the 33 $\mu$m capacitor) is associated to a $\sim$8 \% standard deviation. Other non-vacuum measurements are in the range from 11 \% (33 $\mu$m) to higher values of $\sim$30 \% (50 $\mu$m). The vacuum appears to have stabilized the effect from 12 \% ( $\sigma=1.2*10^{-4}$ N at mean of $1.0*10^{-3}$ N) down to 8 \% ($\sigma=8.7*10^{-5}$ N at mean of $1.1*10^{-3}$ N). 
\begin{figure}[H]
\centering
\includegraphics[scale=1.2]{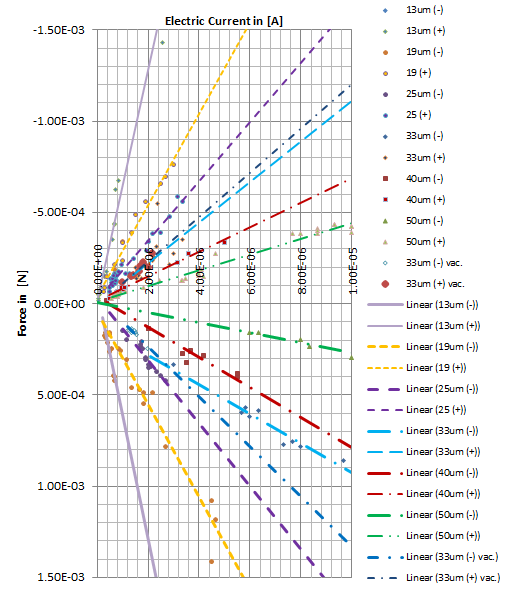}
\small{\caption{Force versus Current}}
\end{figure}

The operational characteristic of the identified thrust effect shows a linear dependency with the amount of accelerated electron mass and increases exponentially with linear decrease of electrode distance.
\\

{\bf{Note:}} The graph shows that for thicker insulators the observed thrust was greater when the anode $(+)$ was oriented on the top of the capacitor. This fact could be correlated with a slight influence of buoyancy determined by the preheating which is required to obtain higher force values.
\\

As advised by Prof. Dr. M. Tajmar, the capacitor plates were tested in vertical position in the attempt to have a null effect that would confirm the absence of external perturbations. This verification test was conducted obtaining a null force (instead of diffuse force directions), confirming the validity of the experimental set-up and the directional attribute of the thrust.
\\

The graph hereunder shows similar trend lines compared to the estimated power consumption.
\begin{figure}[H]
\centering
\includegraphics[scale=1.0]{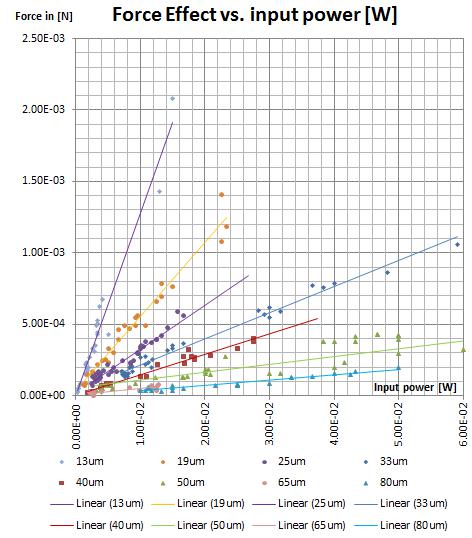}
\small{\caption{Force versus input power}}
\end{figure}
The same graph can be extrapolated to show the force versus power ratio of a scaled-up thruster device up to $1$ kW input power.
\begin{figure}[H]
\centering
\includegraphics[scale=1.0]{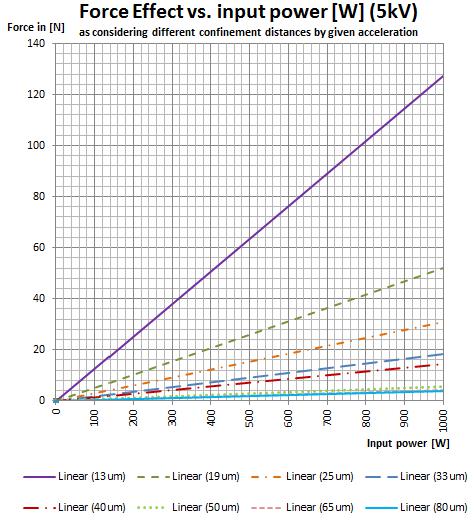}
\small{\caption{Force versus input power extrapolated}}
\end{figure}
In addition, it was found that independent from the electrodes distance the data points line up in a log-graph. The graph highlights how lower thrust performance on thicker insulators is associated with a higher standard deviation (as visible in the lower left corner of the graph) while higher force values, with their lower standard deviation, are obtained with pre-heating in combination with shorter electrodes distance. 

\begin{figure}[H]
\centering
\includegraphics[scale=1.2]{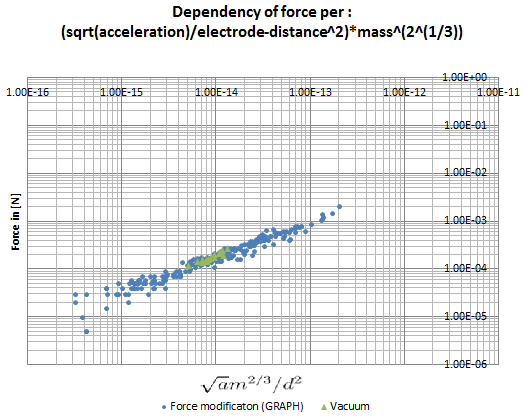}
\small{\caption{Force versus accelerated mass}}
\end{figure}

\subsection{Force versus acceleration voltage for double layer capacitor}

As increased voltage was applied, a slight decrease in normalized force occurred. Voltage was swept from $1$ kV to approximately $10$ kV and a total of $97$ data points were collected. 

\begin{figure}[H]
\centering
\includegraphics[scale=1.2]{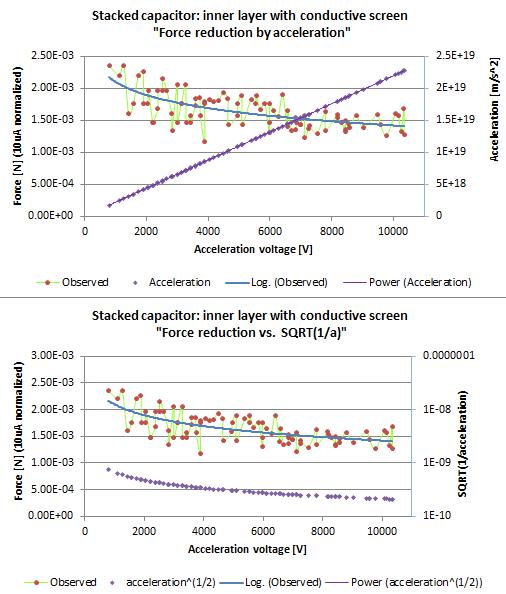}
\small{\caption{Stacked capacitors versus acceleration voltage}}
\end{figure}

Furthermore, capacitor elements were stacked in series in the attempt to obtain a force multiplier factor. With respect to the reduction of the force with an increase in acceleration voltage, it is evident that an increase of the effect corresponds to the reciprocal of the kinetic energy of the accelerated electrons (the effect is reduced with increasing kinetic energy of the particles).
\\

With respect to the force multiplication for these measurements, instead, the preference was to use a modified capacitor which incorporated an additional inner conductive floating material similar to the cathode/anode foil between the two insulators. Such an arrangement represents a series capacitor meaning that the voltage was approximately halved. Identifying that the supply current (see Fig. 14) yielded a comparable equivalent volume of electrons in both the single and double layer capacitor, the results also suggest that the electrons got accelerated twice. It appears that the stacked configuration provides two consecutive accelerations of the particles which have a tangible influence on the output force ($\sim$doubling).

\begin{figure}[H]
\centering
\includegraphics[scale=.4]{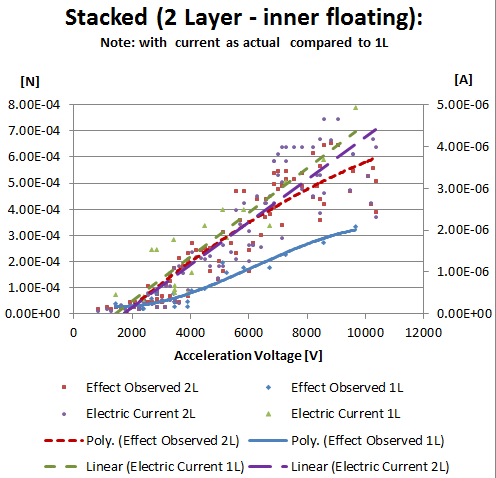}
\small{\caption{Force versus acceleration voltage}}
\end{figure}
\begin{figure}[H]
\centering
\includegraphics[scale=1.4]{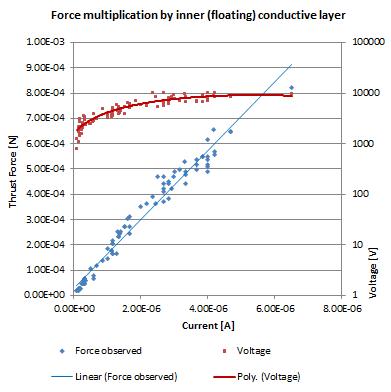}
\small{\caption{Force versus current}}
\end{figure}
\pagebreak

Fig. 16 illustrates the overall weight modification which was observed during the various experiments.

\begin{figure}[H]
\centering
\includegraphics[scale=1.3]{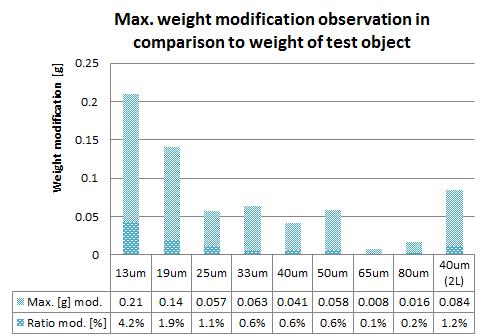}
\small{\caption{Maximum weight modification for all capacitors}}
\end{figure}

An interesting piece of evidence was that some capacitors showed a lower performance in the number of accelerated particles: such performance was detected by measuring a lower electric current. Fig. 16 has the main purpose of illustrating the highest amount of force achieved with different electrode distances.

\subsection{Altering conditions by inserting a conductive wave attenuator around the test object}

Additional tests ($144$ data points) resulted in the identification of a reversal in directional force. Tests had been conducted to determine whether external contributors might have influenced the phenomena. Electromagnetic radiation conducting material had been utilized to dampen radiations with a wavelength lower than a centimeter. 
\\

Below is the simplified skin depth equation for good conductors.
\\
\begin{equation}
\delta=\sqrt{\frac{2\rho}{2\pi f\mu_r\mu_0}}
\end{equation}
\\
$\rho=$ resistivity of the conductor

$f=$ frequency of current

$\mu_r=$ relative magnetic permeability of the conductor

$\mu_0=$ the permeability of free space
\pagebreak

Four different behaviors were identified:
\\

\quad{1. When the thickness of the cathode was extended before a certain range, the thrust effect appeared to be linearly reduced (slope of $-2R$) with respect to the increased elongation of the cathode.}
\\

\quad{{2. When a conductive material, electrically insulated from the cathode, was inserted into a certain area behind the cathode but before a certain range, a full force reversal (compared to the original observed effect) was observed.}
\\

\quad{{3. When a conductive material, electrically insulated from the cathode, was inserted a sufficient distance behind the cathode beyond a range, a normal original effect was observed.}
\\

\quad{{4. When the thickness of the cathode was extended beyond a certain range, a full force reversal (compared to the original observed effect) was observed.}
\\

Suspecting that the forces observed are somehow linked to an external radiation field experienced by the capacitor (virtual particle oscillations, vacuum quantum mechanical effect etc.) during the electron's acceleration, several additional tests were performed in the attempt to dampen shorter wavelengths thus influencing the generated force. This was done introducing an attenuation material, aluminum sheet on the order of a few tenths of a millimeter, at several distances from the surface of the cathode.
\\

After numerous tests, it was discovered that the force could be reversed with approximately the same value of the original force at all distances less than the range, $R$, when the attenuator material was electrically insulated from the cathode. Speculatively, This range might correspond to the rindler horizon equation (which is equal to $c^2/a$) \cite{rr} since the electron experiences high acceleration and this corresponds to the range length.   When placing the attenuator past the range distance $R$ , this led to a sudden change of force direction. This change in direction, with the attenuator sheet beyond the determined range, led to the sudden recovery of the original force direction (towards anode equal to electron propagation path). The distance $R$ within this paper is called “range” or simply $R$ in the remaining part of the paper.
\\

Furthermore it was observed that an extension in thickness of the conducting cathode does not lead to a force reversal but to a reduction of the force. This was estimated as being proportional to the covered space between original position of the cathode surface and the distance which had been identified as $R$.  It should also be noted that there could be an exponential decrease since data points between $0.5R$ and $R$ were not in the range of the measurement device. Finally, if the cathode thickness exceeded $R$ a force reversal was established with the same magnitude of the full normal effect (propagation toward the anode).
\\

The following graph shows coverage condition by the conductive material (insulated and galvanic connected to the cathode) considering previous defined length $R$.
\\

Remark: The force reversal only occurred by covering areas around the cathode. No alteration in the original effect force was observable in other regions (such as in front of the anode).

\begin{figure}[H]
\centering
\includegraphics[scale=0.4]{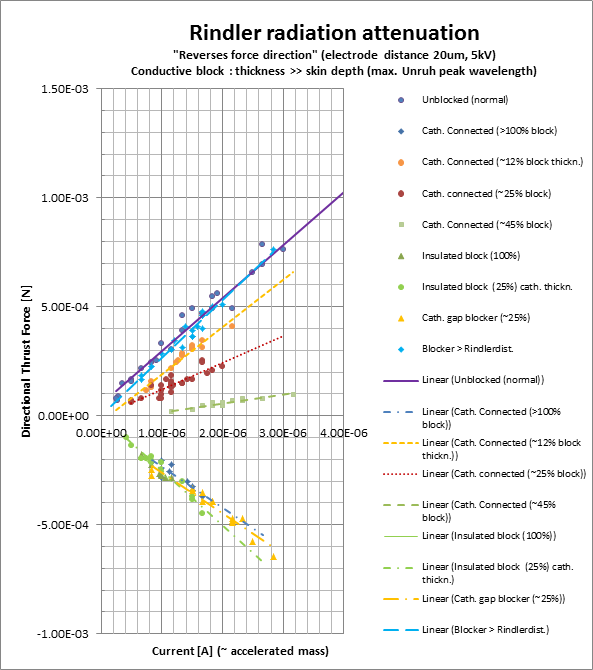}
\small{\caption{Force [N] versus current for various blockers}}
\end{figure}
\begin{figure}[H]
\centering
\includegraphics[scale=1]{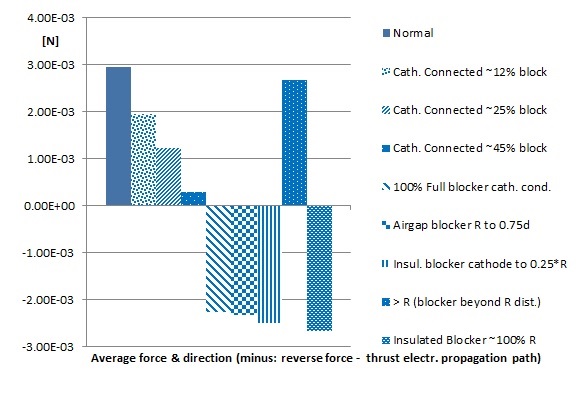}
\small{\caption{Force [N] for different blockers with data measurements averaged}}
\end{figure}
\begin{figure}[H]
\centering
\includegraphics[scale=0.7]{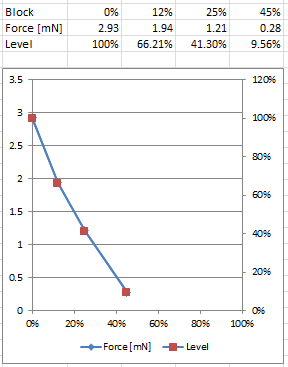}
\small{\caption{Force [N] reduction with blockers with data measurements averaged}}
\end{figure}
\begin{figure}[H]
\centering
\includegraphics[scale=1]{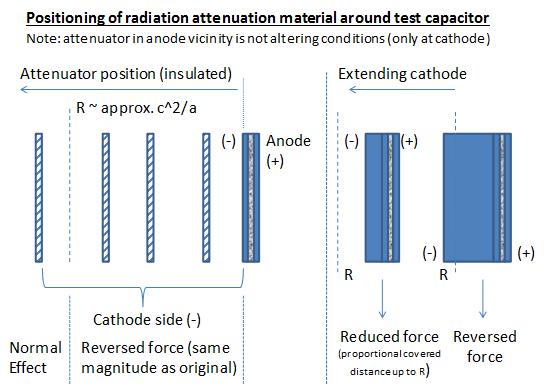}
\small{\caption{Strategic positioning of radiation attenuation materials}}
\end{figure}

\subsection{Wireless Experimental Results}

The new battery-powered onboard wireless setup resulted in a convergent force trendline with the previous data collected for 20 um distance (October 2018). Both devices had a supplied voltage of 5 kV. Generally, smaller thrust values from lower current values were found due to the lower ambient temperature conditions and lack of heat applied to the imFaB device. Artificial heating to increase electric current was avoided to prevent heat trapped inside the closed grounded Faraday cage . This could have resulted in buoyancy errors.
\begin{figure}[H]
\centering
\includegraphics[scale=0.65]{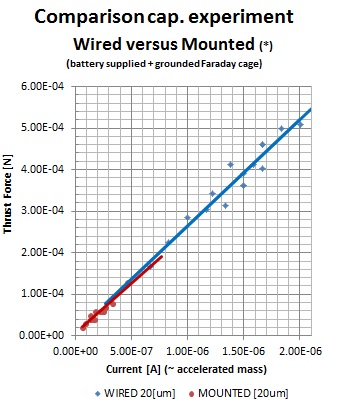}
\small{\caption{Wireless battery powered results}}
\end{figure}

\section{Discussion}

While conducting a variety of capacitive discharge experiments from 2017-2019, a directional force was repeatedly observed during field emission and insulation breakdown discharge of parallel capacitive charged plates. 
\\

The collected experimental data highlighted:
\\

\quad{1. A linear correlation between the thrust force and the accelerated mass.}
\\

\quad{2. An exponential increase in the observed force with the decrease of the capacitor electrode distance, while keeping the accelerated mass constant.}
\\

The effect was observable and repeatable under certain conditions:
\\

\quad{1. With a very short capacitor plate separation distance.}
\\

\quad{2. Under a uniform discharge causing the acceleration of only electrons (as the charged accelerated particles).}
\\

It appears that if an electric discharge occurs, such as bridging the electrodes by arcing, this would introduce positive charges to the system which would hinder the effect. Additionally, the force appears reversible when conductive material, insulated from the cathode, is inserted right behind the cathode of the capacitor, thus behind the accelerated electrons. Additionally, placement of the conductive material below a range, $R$, also changes certain behaviors of directional force which depends on if the material is connected to the cathode or not. Finally, extending the cathode thickness between $R$ and the cathode can also decrease the normal effect. Similar testing was conducted adding conductive material in front of the anode. This led to the same results as without the attenuation material (equal to original effect).
\\

Interestingly, the force appears to correlate to the identified range $R$, as relevant for the force reversal mechanism, within the zone from cathode towards $R$. Here the observed force is plotted against the range $R$.

\begin{figure}[H]
\centering
\includegraphics[scale=1.3]{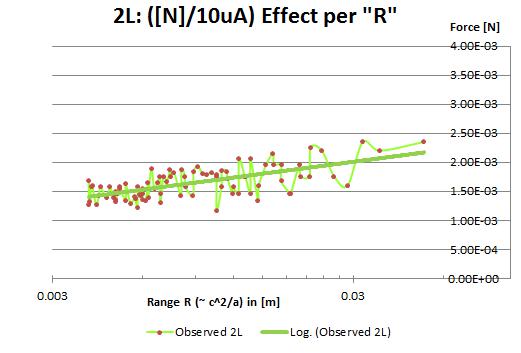}
\small{\caption{Force versus range distance at 10$\mu$A}}
\end{figure}

Overall the illustrated results of the four different sub-experiments, with an adequately high sample size ($507$ data points collected) with evident trend lines, suggest that a new phenomenon, and not an artifact, has been observed. To confirm the evidence, the plan should be to repeat the experiments with a more sophisticated testing facility and equipment. This should provide new evidence and confirmation with a new paper.
\\

It is also worth mentioning that the observed phenomenon does not seem correlated to any evidence found while testing capacitive ion lifters \cite{kk}. The “thrust anomaly”, as described in this paper, is assumed to be caused by electrons accelerated within a capacitor due to a field emission from a high electric field. In this experiment, the capacitor has been used mainly as the mechanism to accelerate particles, in particular, electron charges. Additionally, artifacts are considered to be of a low probability due to the merging trend lines with different experimental setups. This would exclude the fact that two different artifact disturbances could provide data points on the same curve for the device. 
\\

The observations suggest that a wave field around the capacitor is the cause of the thrust and that this field is correlated to accelerated electrons \cite{ss}. This also may suggest that this phenomenon could be a quantum mechanical effect or some relation to Unruh radiation \cite{mm}. This potentially could be confirmed with an existing theoretical framework which seems to predict the observed forces. This additional study, already on-going in 2019, is beyond the purpose of this paper. If preliminary evidence is confirmed, it would open the possibility to design and build fuel-less thrusters that could operate in vacuum, thus suitable for space applications.
\\

Speculatively, in case a Rindler horizon is relevant to the effect, one could consider the inside zone between the cathode and anode to be correlated to the conventional Casimir effect. An electron immersed in this quantum state environment, having an acceleration and by that a Rindler horizon, would be under the influence of an additional field which corresponds to the radiation emitted by the Rindler horizon (reference to the initial hypothesis of Unruh radiation). Hence, the electron might be under the influence of two overlapping fields of quantum states. This would provide an energy/momentum gradient in the fields which, by conservation of momentum law, would initiate a resulting force effect due to the natural symmetry breaking by altering the normal natural radiation which is causing inertia.  The electron floats inside a casimir scenario in between the electrodes. Here, the radiation difference around the electron is not homogenous compared to the normal Casimir effect. Rather the addition of the Rindler horizon radiation causes it to become inhomogeneous.
\\

The new battery-powered onboard wireless setup resulted in successful for bi-directional thrust. A thrust to approximately 10 mg was commonplace for our 20um capacitor with shunt currents matching the wired setup. Earth’s magnetic field which was around 45uT (measured) and an overestimate max length of 0.5 m for supply wires with a draw current of 0.5A in the primary circuit (measured and specs).  Additionally, the wires ran vertically and twisted so would unlikely result in any appreciable Lorentz forces.  Even in a theoretical maximum scenario the influence would be in the 1 mg range and should not affect the overall plausibility of the results. Additionally, a large magnet was placed in the vicinity of the DC wires and there was no noticeable effect during operations. Seems magnetic fields do not have much of an impact as expected for the vertically twisted wires on the scale values. Furthermore, the capacitor device direction was flipped without any change to DC circuit wiring so therefore any Lorentz forces would be always in one direction.
\\

In regards to a prototype thruster (quantum vacuum thruster), the evidence collected would support the claim that the construction of a modular capacitor system, scalable and with relatively low power consumption, could be indeed feasible. The prototypes in this paper have a performance above 0.4 N/kW. This is a performance that Dr. H. White (NASA’s SSRMS Subsystem Manager) considers as a minimum requirement for a crewed mission to Titan/Enceladus \cite{oo}. Construction of modular stacked segments could also provide the advantage of individual segment shut-down in case of failure/malfunction without compromising the thrust performance of the remaining part of the capacitors. Since the thruster concept is exclusively electric, this experimental discovery could provide a first tangible mean for interstellar space exploration once an adequate source of energy is fine-tuned (see performance chart Fig. 11).

\section{Conclusion}

By conducting tests on capacitive systems which accelerate electrons at values on the order of magnitude of $10^{19}$ [m s$^{-2}$] during field emission, a force accelerating the overall system has been clearly identified and characterized. This thrust force, observed distinctively in capacitors with a minimized distance between the electrodes (as a mechanism for particle acceleration), is in linear correlation with the amount of accelerated matter, which corresponds to the electrons released through an electric current determined by field emission. The force is oriented in the propagation direction of the electrons, but can be reversed or attenuated placing conductive material at specific distances in the area behind the cathode (thus outside the capacitor). 
\\

Experimental tests show the force increases exponentially when distance between electrodes is decreased. Also, the force detection was confirmed in soft vacuum condition showing evidence of lower standard deviation of the collected data.
\\

The observed thrust can be enhanced by preheating the capacitor so that the energy value required by the electric field for electron transmission can be lowered by means of Schottky/thermionic effects. In addition, the field strength can be increased by changing the field type from homogeneous to inhomogeneous. For instance using sandpaper or applying a high number of small cuts on the cathode surface can facilitate field emission. Furthermore, the force can be multiplied by using a modular design of stacking capacitors in series: this architecture, with respect to a single, higher performing capacitor, has the advantage of keeping each segment performance independent from the remaining ones.
\\

The capacitors tested in the experiments have a performance above 0.4 [N/kW]; a benchmark often used by NASA to define a thrust ratio sufficient for interplanetary travel. This simple technology, in fact, has the advantage of being completely electric, thus suitable for fuel-less electric propulsion in vacuum. Moreover the scalability of this architecture, coupled with adequate energy source generation, might be appropriate for interstellar travel and precise maneuvering in space. 
\\

Looking into the trend results of the conducted experiments, it should be seen essential to conduct the experiment with more accurate calibrated measuring equipment. It is also suggested that next scientific efforts should focus on correlating these results with an existing theoretical framework, so that an apparent anomaly can be adequately predicted and controlled for practical engineering purposes. The preliminary correlation results are certainly encouraging. A second priority would be to test and/or prepare this apparatus in space to validate its performance in a vacuum and its performance away from a strong gravity source.

\section{Acknowledgments}
We would like to express thanks to Prof. Dr. Martin Tajmar for his advice on the effect verification strategies. We would also like to thank Prof. Dr. M.E. McCulloch for encouraging us and providing us directions \cite{rr}.  Also, we would like to express our gratitude for the editorial support by Mr. Fabio Zagami. Finally, we would like to give thanks to Tommy Callaway for his initial work on our experimental setups. 

\section{Appendix: Potential Errors}

The results during both the remote (battery powered unit without exposed HV conductors) and wire testing could result in “Trichel pulses" by insufficient insulation. The signature of these pulses have the properties of lifted currents (offset = DC component with sharp inrush currents) as seen on the oscilloscope display. The signals look similar to field emission currents but have different attributes (spikes), and are actually a corona discharge with trichel pulses. Field emission events occur having current peaks starting from approximately zero on the display of the oscilloscope. In contrast, external located trichel impulses (in wiring etc.), which are generated outside the test capacitor, do fluctuate but are added in amplitude to the small DC component leakage current. It was found that a leakage current is likely generated from a conductive enclosure such as the Faraday shielding cage without sufficient inner insulation material near a thin high voltage charged conductor. Here, it would also be relevant to check the insulation rating when appropriate since, for example, a 500 V rating would be insufficient as a direct barrier between HV polarities. Usually, 500 V is a typical standard value rating of insulators on the market. Furthermore, an air layer with the insulation material of a conductor could become a composite capacitor (voltages distribute with each material index and per thickness involved). The electric field strength inside the air is significantly higher than inside the material. Therefore if a metallic object, such as a shielding enclosure, is connected to the negative polarity, electrons could be tunneled out the surface due to insufficient insulation and ionize the air gap to the next conductor. Additionally, using conductors not rated appropriately to the intended use could also lead to a null effect due to leakage currents. Additionally, having twisted conductor supplies inside sheeted routing should not be near the enclosure which could have a different polarity. If the field is strong enough the electrons could be pulled out of the enclosure and accelerated towards the insulator. 
\\

One can also consider the influence of possible Lorentz forces even thought the HV supply current is significantly low. One solution is the usage of twisted conductors. However, this should be done with care as twisted conductors with different polarities with longer lengths could also cause leakages. Hence it should be seen essential to provide twisting with limited/avoiding direct contact of insulation of conductors with different polarities during a high voltage scenario. During twisting, application of air loops may boost insulation distance and limit the direct contact points.  Also, it is necessary to utilize conductors rated for the high voltage application.  Adding tubing over high voltage conductors rated up to 500V (working voltage, insulation voltage) should be considered an insufficient insulation system if there is direct contact with one conductor (Faraday or conductive mu-metal surface connected to one HV polarity). Recreation of the Trichel pulses during were performed in January 2019 during further testing. The result usually ended in a null effect since the path of least resistance was not through the capacitor. Larger currents around 10uA were commonplace measured at the shunt. For a valid effect, values were nominally much lower in the low uA range for a large normal effect.
\\

At the current state of research it is not yet clarified whether the quantum vacuum oscillations emerge from the local vacuum or are emitted by the cosmic horizon \cite{rr}. A mu-metal shielding might cause complications during testing so in the future it is suggested to compare the use of normal Faraday cages and mu-metal cages.

\bibliographystyle{abbrv}
\bibliography{arxivfabnew}

\end{document}